\documentclass[12pt]{article}
\usepackage{a4}
\usepackage{psfig}
\usepackage{epsfig}
\usepackage{epsf}
\usepackage{rotating}
\usepackage{citesort}
\usepackage{amssymb}
\usepackage{graphicx}
\newcommand{\lsim}{\mbox{$\leq$}}

\newcommand{\anueR}{\mbox{$\overline{\nu}_{eR}$}}

\newcommand{\anureaction}{\mbox{$\overline{\nu}_e+p\to n+e^{+}$}}

\def\plb#1#2#3{    { Phys. Lett. }{\bf B #1} (19#2) #3}

\begin{document}
\phantom{peppe}
\vspace{-2cm}
\begin{flushright}
{\small
IFUM-700/FT\\
FTUAM-02-457\\
}
\end{flushright}
\begin{center}
\vspace{0.5cm}
{ \Large \bf KamLAND, solar antineutrinos and their magnetic moment}\\[0.2cm]
\large P.~Aliani$^{a\star}$, V.~Antonelli$^{a\star}$, M.~Picariello$^{a\star}$, E.~Torrente-Lujan$^{ab\star}$\\[2mm]
$^a$ {\small\sl Dip. di Fisica, Univ. di Milano},
{\small\sl and INFN Sez. Milano,  Via Celoria 16, Milano, Italy}\\
$^b$ {\small\sl Dept. Fisica Teorica C-XI, 
Univ. Autonoma de Madrid, 28049 Madrid, Spain}\\
\end{center}

\begin{abstract}
We investigate  the possibility of detecting 
solar antineutrinos with the KamLAND experiment. 
These antineutrinos  are 
predicted by spin-flavor oscillations at a significant rate 
even if this mechanism is not the leading solution to the 
SNP. The recent evidence from SNO shows 
that a) the neutrino oscillates, only around 34\% of the 
initial solar neutrinos arrive at the Earth as electron 
neutrinos and b) the conversion is mainly into active 
neutrinos, however a non e,$\mu,\tau$ component is allowed: 
the fraction of oscillation into  non-$\mu-\tau$ neutrinos 
is found to be $\cos^2\alpha= 0.08^{+0.20}_{-0.40}$. This 
residual flux could include sterile neutrinos and/or  the 
antineutrinos of the active flavors.

KamLAND is potentially sensitive to antineutrinos derived 
from solar ${}^8$B neutrinos.
In case of negative results, we find that  KamLAND could put 
strict limits on the flux of solar antineutrinos, 
$\Phi( {}^8 B)< 1.0\times 10^4\ cm^{-2}\ s^{-1}$,
 more than one order of magnitude smaller than existing
 limits,
and on their appearance probability 
$P<0.20-0.15\%$ (95\% CL)
after 1-3 years of operation. Assuming a concrete model for 
antineutrino production by spin-flavor precession,
 this upper bound   implies an 
upper limit on the product of the intrinsic neutrino magnetic 
moment and the value of the solar magnetic
  field $\mu B< 10^{-21}$ MeV (95\% CL).
For $B\sim 10-100$ kG, 
we would have  $\mu <  10^{-11}-10^{-12}\ \mu_B$ (95\% CL).

In the opposite   case, if spin-flavor precession 
is indeed at work even at a non-leading rate, the additional 
flux of antineutrinos could strongly distort the signal 
spectrum seen at KamLAND at energies above 4-5 MeV and 
their contribution should properly be taken into account.

\vskip .5truecm

{PACS: }

\vfill
{\small \noindent 
$\star$ email: paul@lcm.mi.infn.it, vito.antonelli@mi.infn.it,
 marco.picariello@mi.infn.it, torrente@cern.ch}
\end{abstract}

\newpage


\section{Introduction}

The publication of the recent SNO results~\cite{Ahmad:2002ka,Ahmad:2002jz,howto}
has made an important breakthrough towards the solution of the long standing
 solar neutrino 
problem (SNP) possible
\cite{Aliani:2002ma,Strumia:2002rv,Bandyopadhyay:2002xj,Barger:2002iv,Pascoli:2002xq,deHolanda:2002pp,Bahcall:2002hv,Foot:2002re,Aliani:2002er}.
These results provide the strongest evidence so far 
for flavor oscillation in the neutral lepton sector. 
However the concrete mechanism behind neutrino oscillations is far from 
being dilucidated. Spin Flavor  oscillations 
\cite{generalrandom,specificrandom} 
could be at work and constitute, 
if not the main ingredient of the 
explanation of the SNO results and the Solar Neutrino problem, at 
least a sub-leading ingredient.

In the near future the reactor experiment KamLAND~\cite{piepke,kamLANDmonaco} 
is expected to further improve our knowledge of neutrino mixing. In fact it 
should be able to sound the region of the mixing parameter space corresponding
to the so called Large Mixing  Angle (LMA) solution of the solar neutrino 
problem  ($\Delta m^2\sim 10^{-5} - 10^{-4} \ eV^2$ and $\tan^2\theta\sim 10^{-1} - 1 $)
more profoundly.

The previous generation of reactor experiments 
(CHOOZ~\cite{CHOOZ}, PaloVerde~\cite{PaloVerde}), performed with a baseline 
of about 1 km. They have attained  a sensitivity of 
$\Delta m^2<10^{-3}\ eV^2$ \cite{chooznew,CHOOZ} and, not finding any 
dissapearence of the initial flux, they demonstrated that the atmospheric 
neutrino anomaly \cite{atmospheric} is not due to muon-electron neutrino 
oscillations. The KamLAND experiment is the successor of such experiments at 
a much larger scale in terms of baseline distance and total incident flux.  

This experiment relies upon a 1 kton liquid scintillator detector located at 
the old, enlarged,  Kamiokande site. It searches for the oscillation of 
antineutrinos emitted by several nuclear power plants in Japan. The nearby 
16  (of a total of 51) nuclear power stations deliver a $\overline{\nu}_e$ 
flux of $1.3\times 10^6 cm^{-2}s^{-1}$ for neutrino energies $E_\nu>1.8$ MeV 
at the detector position. About $78\%$ of this flux comes from 6 reactors 
forming a well defined baseline of 139-214 km. Thus, the flight range is limited 
 in spite of using  several reactors, because of this fact the sensitivity 
of KamLAND will  increase by nearly two orders of magnitude compared to previous 
reactor experiments.

Beyond reactor neutrino measurements, the planned secondary physics program 
of KamLAND includes diverse objectives as the measurement of geoneutrino flux 
emitted by the radioactivity  of the earth's crust and mantle, the detection 
of antineutrino bursts from galactic supernova and, after extensive  
improvement of the detection sensitivity, the detection of low energy 
${}^7 Be$ neutrinos using neutrino-electron elastic scattering.

In this work we want to stress another possibility. 
The KamLAND experiment is potentially capable of 
detecting  antineutrinos produced on fly 
from solar ${}^8$B neutrinos.
These  antineutrinos are predicted by spin-flavor oscillations 
at a significant rate if the neutrino is a Majorana particle and 
if its magnetic moment is high enough
\cite{generalrandom,specificrandom}.  
Let us remark that the flux of reactor antineutrinos at the Kamiokande site 
is comparable, and in fact smaller, to the flux of 
${}^8$B neutrinos emitted by the sun
($\Phi( {}^8 B)\simeq 5.6\times 10^6 cm^{-2}s^{-1}$, 
\cite{bpb2001,Ahmad:2002jz}).
Their energy spectrum peaks at a somehow lower point, a detail which is important
 for their detection as a way of separating them from reactor antineutrinos.
This is graphically shown in Fig.(1).

Let us briefly recall some  model independent conclusions
obtained from the  results of  
SNO~\cite{Aliani:2002ma,Barger:2001zs}.
From the three fluxes measured by SNO and from
 the  flux predicted by the solar standard 
mode one can define, 
following   Ref.\cite{Barger:2001zs},  the 
quantity $\sin^2\alpha$, 
one finds~\cite{Aliani:2002ma}.
$$
\sin^2\alpha=0.92^{+0.39}_{-0.20},
$$
where the SSM flux is taken as the ${}^8\rm B$ flux predicted 
in Ref.\cite{bpb2001}.
The central value is clearly below one (only-active oscillations).
The same can be  written in another way, 
the fraction of oscillating 
neutrinos into non-active ones is
$$\cos^2\alpha=0.08_{-0.39}^{+0.20}.$$
As a conclusion from these numbers,
the hypothesis of transitions to {\em only} sterile 
 neutrinos is rejected at nearly $5\sigma$, however
electron neutrinos are still {\em allowed}
 to oscillate into sterile neutrinos

\begin{figure}
\centering
\begin{tabular}{c}
\psfig{file=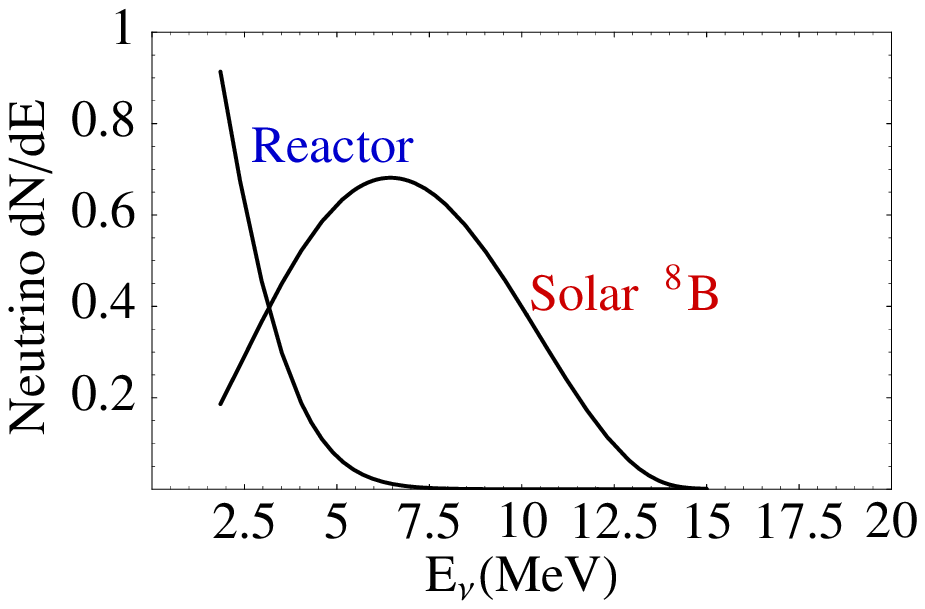,width=9.5cm} \\ 
\psfig{file=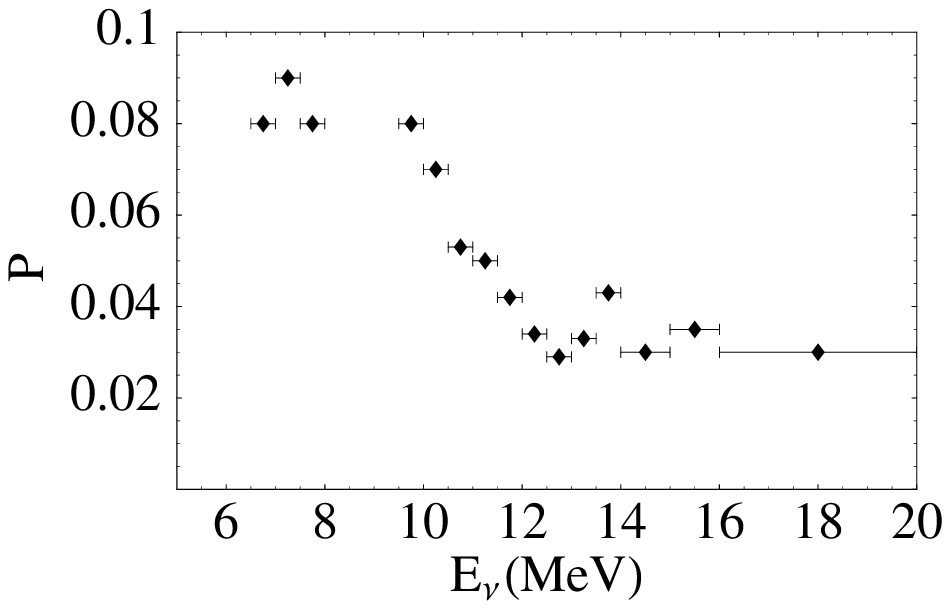,width=9.5cm}
\end{tabular} 
\caption{(Top)
The reactor antineutrino  and solar ${}^8 B$ neutrino 
\protect\cite{bpb2001} fluxes.
(Bottom).
Upper limits on solar antineutrino conversion probabilities 
(from Ref.\protect\cite{antibounds}).
}
\label{figurebounds}
\end{figure}

In terms of absolute flux, the values above for $\cos^2\alpha$ means that 
the non-standard flux and a fortiori, the solar antineutrino 
flux is limited below 
$\Phi(\overline{\nu}_{sun})\sim 1\times 10^6$ cm$^{-2}$ s$^{-1}$.
The existing  bounds on solar antineutrinos are however much stricter.
 The present upper limit on the absolute flux of solar antineutrinos
originated from ${}^8 B$ neutrinos is \cite{antibounds,PDG2002}
$\Phi_{\overline{\nu}}({}^8 B)< 1.8\times 10^5\ cm^{-2}\ s^{-1}$
which is equivalent to an averaged conversion probability bound
of $P<3.5\%$ (SSM-BP98 model). There are also bounds on their differential 
energy spectrum \cite{antibounds}: the conversion probability is smaller 
than $8\%$ for all $E_{e,vis}>6.5$ MeV going down the $5\%$ level above 
$E_{e,vis}\simeq 10$ MeV, these results are 
summarized in Fig.\ref{figurebounds}.

\section{A KamLAND overview}
\label{kamland}

Independently of their origin, solar or reactor
electron antineutrinos from nuclear reactors  with energies 
above 1.8 MeV can be detected in  KamLAND by  the inverse 
$\beta$-decay reaction $\overline{\nu}_e+p\to n+e^+$. The time 
coincidence, the space correlation and the energy balance  
between the positron signal and the 2.2 MeV $\gamma$-ray
 produced by the capture of a already-thermalized  neutron on a
 free proton make it possible to identify this reaction 
unambiguously, even in the presence of a rather large background. 

The main  ingredients in the calculation of the 
corresponding expected 
signals in KamLAND are  solar fluxes mentioned above, the reactor  flux and
the antineutrino cross section on protons. These last two 
 are considered below (see also Ref.\cite{Aliani:2002ca}). 

\subsection{The reactor antineutrino  flux}

We first describe the flux of antineutrinos coming from 
the power reactors.
A number of short baseline experiments 
(Ref.\cite{Murayama:2000iq} and references therein) 
have measured the energy spectrum of reactors at distances 
where oscillatory effects have been shown to be inexistent. 
They have shown that the theoretical neutrino flux predictions 
are reliable within 2\% \cite{piepke}.

The effective flux of antineutrinos released by the nuclear
 plants is a rather well  understood function of 
the thermal power of the reactor and
 the amount of thermal power emitted during the 
fission of a given nucleus, which gives the total amount, and 
the  isotopic composition of the reactor fuel which gives the 
spectral shape.
Detailed tables for these
 magnitudes can be found in Ref.~\cite{Murayama:2000iq}.

For a given isotope $(j)$ the energy  spectrum can be parametrized 
by the following expression 
$d N_\nu^{j}/d E_\nu=\exp (a_0+a_1 E_\nu+ a_2 E_\nu^2)$
where the coefficients $a_i$ depend 
on the nature of the fissionable isotope 
(see Ref.\cite{Murayama:2000iq} for explicit values).
Along the year, between periods of refueling, the total effective flux changes with time as the fuel is expended and the isotope 
relative composition varies.
The overall spectrum is at a given time
$ {d N_\nu}/{d E_\nu}=\sum_{j=isotopes} 
c_j(t){d N_\nu^{j}}/{d E_\nu}.$
To compute a fuel-cycle averaged spectrum
we have made use of the typical time 
evolution of the relative 
abundances $c_j$, which  can be seen in Fig. 2 of 
Ref.\cite{Murayama:2000iq}.
This averaged spectrum can  be again  fitted very well by 
the same functional expression as above.
The isotopic energy yield  is properly taken into account. 
As the result of this fit, we obtain 
the following values which are the ones to be used in the 
rest of this work:
$a_0=0.916,\quad a_1=-0.202,\quad a_2=-0.088.$   
Although individual variations of the $c_j$ along the 
fuel cycle can be very high, the variation of the two most 
important ones is highly correlated: the 
coefficient $c({}^{235} U)$ increases in the range
 $\sim 0.5-0.7$ while 
$c({}^{239} Pu)$ decreases $\sim 0.4-0.2$. 
This correlation makes  the effective description of the 
total spectrum by a single expression as above useful.
With the fitted coefficients $a_i$ above, the difference between
this effective spectrum and the real one is typically $2-4\%$ 
along the yearly fuel cycle.

\subsection{Antineutrino cross sections}

We now consider the
cross sections for antineutrinos on protons. We will 
sketch the form of the well known differential expression and 
more importantly we will give updated numerical values 
for the transition matrix elements which appear as 
coefficients. 

In the limit of infinite nucleon mass, the 
cross section  for the reaction 
\anureaction\  is given by \cite{zacek,reines} 
$\sigma(E_{\overline{\nu}})=k\  E_{e^+} p_{e^+}$
where $E,p$ are the  positron energy and momentum
 and $k$ a  transition matrix element which will be 
considered 
below.
The  positron spectrum  is  monoenergetic: 
 $E_{\overline{\nu}}$ and  
$E_{e^+}$ are related by:
$E_{\overline{\nu}}^{(0)}=E_{e^+}^{(0)}+\Delta M$,
where $M_n, M_p$ are the neutron and proton masses
 and $\Delta M=M_n-M_p\simeq 1.293 $ MeV.

Nucleon recoil corrections are 
potentially important in relating the positron and antineutrino 
energies in order to evaluate the antineutrino flux. 
Because the antineutrino 
flux $\Phi(E_\nu)$ would typically decrease quite 
rapidly with energy, the lack of adequate corrections 
will systematically overestimate the   positron yield.   
For both cases, solar or reactor antineutrinos, 
 because the antineutrino 
flux $\Phi(E_\nu)$ would typically decrease quite 
rapidly with energy, the lack of adequate corrections 
will systematically overestimate the   positron yield.   
For the solar case and taking into account the SSM-BP98 
${}^8 B$ spectrum, the effect  decrease the positron 
yield by 2-8\% at the main visible energy range $\sim 6-10$ MeV.
The positron yield could decrease up 50\% at {\em hep} 
neutrino energies, a region where incertitudes in the total 
and differential spectrum are of comparable size or larger.
 Finite energy resolution smearing will however diminish  
this correction when integrating over large enough energy 
bins: in the range $6.5-20$ MeV the net positron suppression 
is estimated to be at the $5\%$ level, increasing up $20\%$ 
at {\em hep} energies.

At highest orders,  the  positron spectrum  is not 
monoenergetic and one has to integrate over the positron angular 
distribution to obtain the positron yield.
We have used the 
complete expressions which  can be found 
in Ref. \cite{vogel}. Here we only want to stress 
the numerical value of 
the overall coefficient $\sigma_0$ (notation of 
Ref.\cite{vogel}) which is related to the 
transition matrix element $k$ above.
The matrix transition element can be written in terms of 
measurable quantities as 
$k=2\pi^2 \log 2/ (m_e^5 f \, t_{1/2}).$
Where the value of the space factor
  $f=1.71465\pm 0.00015$ follows from calculation 
\cite{zacek21},
 while $t_{1/2}=613.9\pm 0.55$ sec is the latest published value for 
the free neutron half-life \cite{PDG2002}. This value has a  significantly 
smaller  error than previously quoted measurements.
From the values above, we obtain the extremely precise value: 
$k=(9.5305\pm 0.0085)\times 10^{-44}\  cm^2/MeV^{2}.$ 
From here the coefficient which appears in the differential 
cross section is obtained as
(vector and axial vector couplings $f=1,g=1.26$):
$ k=\sigma_0 (f^2+3 g^2).$
In summary, the differential cross section which appear in
  KamLAND are very well known, its theoretical 
errors are negligible if updated values are employed.

\section{The   solar signal and reactor backgrounds}
\label{klsignal}

The average number of positrons $N_i$ originated from the solar source which 
are detected per visible energy bin $\Delta E_i$ is given by the convolution 
of different quantities: 
\begin{eqnarray} \hspace{-0.3cm}
N_i&=& Q_0 \int_{\Delta E_i}dE_e \int_0^\infty dE_e^r \epsilon(E_e)R(E_e,E_e^r)
\int_{E_e^r}^\infty dE_{\overline{\nu}} \overline{P}(E_{\overline{\nu}})
\Phi (E_{\overline{\nu}})
 \sigma (E_{\overline{\nu}},E_e^r)
\label{e3466} 
\end{eqnarray}
where $Q_0$ is a normalization constant  accounting for the 
fiducial volume and live time of the experiment,
$\overline{P}$, the neutrino-antineutrino 
oscillation probability.
the antineutrino capture cross section
$\sigma (E_{\overline{\nu}},E_e^r)$ given as before.
The functions $\epsilon(E_e)$ and $R(E_e,E_e^r)$ are
 the detection efficiency and  the energy 
resolution function.
We suppose  a perfect detector efficiency $\epsilon\sim 1$.
and energy resolution 
$\sigma(E)/E\sim 10\%/\surd E$
\cite{brackeeler,kamLANDmonaco}.
In order to obtain concrete limits, a model should  be taken which 
predict $\overline{P}$ and its dependence with the energy.
For our purpose it will suffice  to suppose  $\overline{P}$ a 
constant over the  Boron solar energy range.


Similarly,  the expected numbers of positron events originated 
from power reactor neutrinos are obtained summing 
the expectations for all the relevant 
reactor sources weighting each source by its power and distance to the detector
(table II in Ref.~\cite{Murayama:2000iq}), assuming the same spectrum originated 
from each reactor. 
We have used the antineutrino flux spectrum  given by  
the expression  of the previous section and  the 
relative reactor-reactor power normalization.

For one year of running with the 600 ton fiducial mass and for standard nuclear 
plant power and fuel schedule: we assume all the reactors operated at 
$\sim 80\%$ of their maximum capacity and an averaged, time-independent, fuel 
composition equal for each detector, the experiment expects about 550 
antineutrino events.

In addition to the  reactor antineutrino  signal deposited in the detector, 
two classes of other backgrounds can be distinguished 
\cite{piepke,Murayama:2000iq,brackeeler}. The so called random coincidence 
background is due to the contamination of the detector scintillator by 
U, Th and Rn. From MC studies and assuming that an adequate level of
purification can be obtained, the background coming from this source
is expected to be $\sim 0.15$ events/d/kt which is 
equivalent to a signal to background ratio of  $\sim 1\%$. 
Other works \cite{usareport} conservatively estimate 
a $5\%$ level for this ratio. More importantly for what it follows, 
one expects that the 
random coincidence  backgrounds will be a relatively steeply 
falling function of energy. The assumption of no random coincidence 
background should be relatively safe at high energies above $\sim 5$ MeV
which are those of interest here.

The second source of background, the so called correlated background is 
dominantly caused by cosmic ray muons and neutrons. The KamLAND's depth is
 the main tool to suppress those backgrounds. MC methods estimate a correlated 
background of around $0.05$ events/day/kt distributed over all the energy 
range up to $\sim 20$ MeV, this is the quantity that we will consider later.


Reactor and solar antineutrino signals are shown in 
Fig.(2) and table (1). These results will be discussed in the
next section.

\section{Results and Discussion}

In order to estimate the sensitivity of KamLAND to 
put limits on the flux of antineutrinos arriving from the 
sun we have computed the expected signals coming from solar and 
reactor  antineutrinos and from the background. 
They  are presented in Table (\ref{table1})
 for different representative 
values of the minimum energy required ($E_{thr}$) for the 
visible positrons. 
We have supposed a background of $0.05$ evt/d/kt uniformly 
 distributed over the full energy range.
To obtain the solar numbers (first column, $S_{sun}$)
 we have supposed full  neutrino-antineutrino conversion  
($\overline{P}=1$) 
with no spectral distortion.
For any other conversion probability, the experiment should see 
the  antineutrino 
quantity $\overline{P}\times S_{sun}$ 
in addition to the reactor ones and other background. 
If the experiment does not receive any solar antineutrinos, making 
a simple statistical estimation (only statistical errors are 
included) we obtain the upper limits on the 
conversion probability which appear in the last column of the table.

From the table we see that 
after three years of data taking the optimal result is obtained imposing a 
energy detection threshold at $\sim 7$ MeV. A negative result would allow 
to impose an upper limit on the average antineutrino appearance probability 
at $\sim 0.20\%$ (95\% CL). The corresponding limits after one year of data 
taking are only  slightly worse, they are respectively: 0.21-0.24\% (95\% CL).

These results are obtained under the supposition of no disappearance on 
the reactor flux arriving to KamLAND. No flux suppression is  expected for 
values of the mixing parameters in the LOW region, more precisely for any
$\Delta m^2 \leq 2 \times 10^{-5}$ eV$^2$ (see Plot 1(right) in 
Ref.\cite{Aliani:2002ma} and Ref.\cite{Aliani:2002ca}).
The consideration of  reactor antineutrino oscillations does not change 
significantly the sensitivity in obtaining upper limits on $\overline{P}$.
For values of the mixing parameters fully on the LMA region,
$\Delta m^2 \geq 1-9 \times 10^{-5}$ eV$^2$, the flux suppression is 
typically $S/S_0\sim 0.5-0.9$ and always over $S/S_0\sim 0.4$, for any 
the energy threshold $E_{thr}\sim 5-8$ MeV.
We have obtained the expected reactor antineutrino contribution  
for a variety of points in the LMA region 
(see table I in Ref.\cite{Aliani:2002ca}) and 
corresponding upper limits on $\overline{P}$: the results after 
3 years of running are practically the same while for 1 year of 
data running are slightly better (for example $\overline{P}$ goes 
down from 0.27 to 0.3 for $E_{thr}>6$ MeV. 

\section{A model for solar antineutrino production}

The combined action of  spin flavor precession in a 
magnetic field and  ordinary neutrino matter oscillations 
can produce an observable flux of $\anueR$'s from the Sun 
in the case of the neutrino being  a Majorana particle.
In the simplest model, where a thin layer of highly chaotic of 
magnetic field is assumed at the bottom of the convective 
zone (situated at $R\sim 0.7 R_\odot$), the antineutrino appearance 
 probability at the exit of the 
layer $P_f$
(and therefore basically 
the appearance probability of antineutrinos 
at the earth)
can be written as \cite{specificrandom}
 (see also Refs.\cite{generalrandom}):
\begin{equation}
P(\nu_{eL}\to  \tilde{\nu}_{eR}  )_f  = \xi\times
 P(\nu_{eL}\to \nu_{\mu L} )_i,
\end{equation}
where $P_f$ is the $e-\mu$ conversion probability at the 
entry of the layer and where
 the constant 
$1-2\xi\simeq\exp(-4 \Omega^2 \Delta r)$ summarizes  the 
effect of the magnetic field. 
This quantity   depends on the layer
width $\Delta r\ (\sim 0.1 R_\odot$), 
the r.m.s strength of the
chaotic field $\Omega^2\ (=\mu^2 L_0 \langle B^2\rangle/3$).
$B$ is the magnitude of the magnetic field and $L_0$ is
 a scale length ($L_0\sim 1000$ km).
For small values of the argument we have 
$\xi\sim 2 \Omega^2 \Delta r$.
The antineutrino flux, $P_f$, could be large if $P_i$ is 
large, i.e., the 
 neutrino have passed through a MSW resonance  before arriving 
to the layer. 
The MSW resonance  converts practically all the 
initial $\overline{\nu}_e$ flux into $\overline{\nu}_{\mu}$. 
The  field finally converts them into $\overline{\nu}_e$. 
A fraction  of the $\overline{\nu}_e$ will be reconverted 
into  $\overline{\nu}_{\mu}$ by mass oscillations but this 
reconversion is limited in this case by the chaotic character 
of the process.

A detailed computation of the expected flux and average  
appearance probability of electron antineutrinos detectable 
at SuperK and SNO has been performed in 
Ref.\cite{specificrandom}. 
The conclusion of this work (see Figs.1 and 2 there) is that 
the appearance probability is above the 1\% level for 
the region of the parameter space allowed by present combined
evidence and for any value of the parameter $\xi$ such 
that $\xi>0.02$.
 Otherwise, if we lower $\xi$ down  $\xi\sim 0.002$, 
the antineutrino appearance probability takes values in 
the range $0.1-1\%$
for all the parameter space allowed by present experiments.
These values are  within the sensitivity of the the KamLAND 
experiment as we have seen in the previous section.

Upper limits on the antineutrino appearance probability 
can be translated into upper limits on the parameter $\xi$
 and then on the neutrino magnetic moment.

In case of  negative finding, KamLAND  will be able to impose 
an upper  bound $P\sim 0.2\%$. We can translate this on an 
upper bound on $\xi$.  
An upper limit $\xi<0.002$  implies an upper limit on 
the product of the intrinsic neutrino magnetic moment and
the value of the convective solar magnetic field as 
$\mu\ B< 10^{-21}$ MeV (95\% CL).
For realistic values of other astrophysical solar
 parameters ($B\sim 15$ kG), these upper 
limits would imply that the neutrino magnetic 
moment is constrained to be $\mu\lsim 10^{-11}\ \mu_B$
(95\% CL).
For $B\sim 10-100$ kG, 
we would have  $\mu <  10^{-11}-10^{-12}\ \mu_B$.

\section{  Conclusions}
\label{sec:conclusions}

In summary in this work we investigate  the possibility of 
detecting 
solar antineutrinos with the KamLAND experiment.
These antineutrinos are predicted by spin-flavor solutions
to the solar neutrino problem.
The recent evidence from SNO  shows that 
a) the neutrino oscillates, only around 34\% of the 
initial solar neutrinos arrive at the Earth as electron 
neutrinos and
b) the conversion is mainly into active neutrinos,
 however a non e,$\mu-\tau$ component is allowed: 
the fraction of oscillation into  non-$\mu-\tau$ neutrinos 
is found to be 
$\cos^2 \alpha= 0.08^{+0.20}_{-0.40}$.
This residual flux could include sterile neutrinos and/or  
the antineutrinos of the active flavors.

The KamLAND experiment is potentially sensitive 
to antineutrinos coming from solar ${}^8$B neutrinos.
In case of negative results, we find that the 
results of the KamLAND experiment could put 
strict limits on the flux of solar antineutrinos
$\Phi( {}^8 B)< 1.0\times 10^4\ cm^{-2}\ s^{-1}$,
and their appearance probability ($P<0.2-0.15\%$), respectively 
after 1-3 years of operation.
Assuming a concrete model for antineutrino production 
by spin-flavor precession in the convective solar 
magnetic field, this upper bound on the 
appearance probability  implies an upper limit on 
the product of the intrinsic neutrino magnetic moment and
the value of the  field $\sim \mu B< 10^{-21}$ MeV.
For $B\sim 10-100$ kG, 
we would have  $\mu <  10^{-11}-10^{-12}\ \mu_B$.

In the opposite   case, if spin-flavor precession 
is indeed at work even at a minor rate, 
the additional flux of 
antineutrinos could strongly distort the 
signal spectrum seen at KamLAND at energies above 
4 MeV and their contribution should be
taken into account. This is graphically shown in 
Fig.(2)

\vspace{0.3cm}
\subsection*{Acknowledgments}
We  acknowledge the  financial  support of 
 the Italian MURST and  the  Spanish CYCIT  funding 
agencies. One of us (E.T.) wish to acknowledge in 
addition the hospitality of the 
CERN Theoretical Division at the early stage of
this work. 
P.A., V.A., and M.P. would like to thank the 
kind hospitality of the Dept. de Fisica Teorica of the
U. Autonoma de Madrid.
The numerical calculations have 
been performed in the computer farm of 
 the Milano University theoretical group.

\newpage

\begin{table}[t]
 \begin{center}
  \scalebox{0.99}{
    \begin{tabular}{lcrccc}
\hline
  E$_{thr}$     & $S_{Sun}$& $S_{Rct}$ & Bckg. &
 P (CL 95)\% & P (CL 99)\% \\
\hline
6 MeV  & 616  & 43 & 70 & 0.22 & 0.23          \\
7 MeV  & 500  & 11 & 65 & 0.19 & 0.20         \\
8 MeV  & 366  & 2 &  60 & 0.21 & 0.23         \\
\hline
      \end{tabular}
}
 \caption{ Expected signals from solar antineutrinos  after 
3 years of data taking. Reactor antineutrino (no oscillation is 
assumed) and other background 
(correlated background) over the same period. The random coincidence 
background is supposed negligble above these energy thresholds. 
Upper limits on the antineutrino oscillation probability.
}
  \label{table1}
 \end{center}
\end{table}

\begin{figure}
\centering
\psfig{file=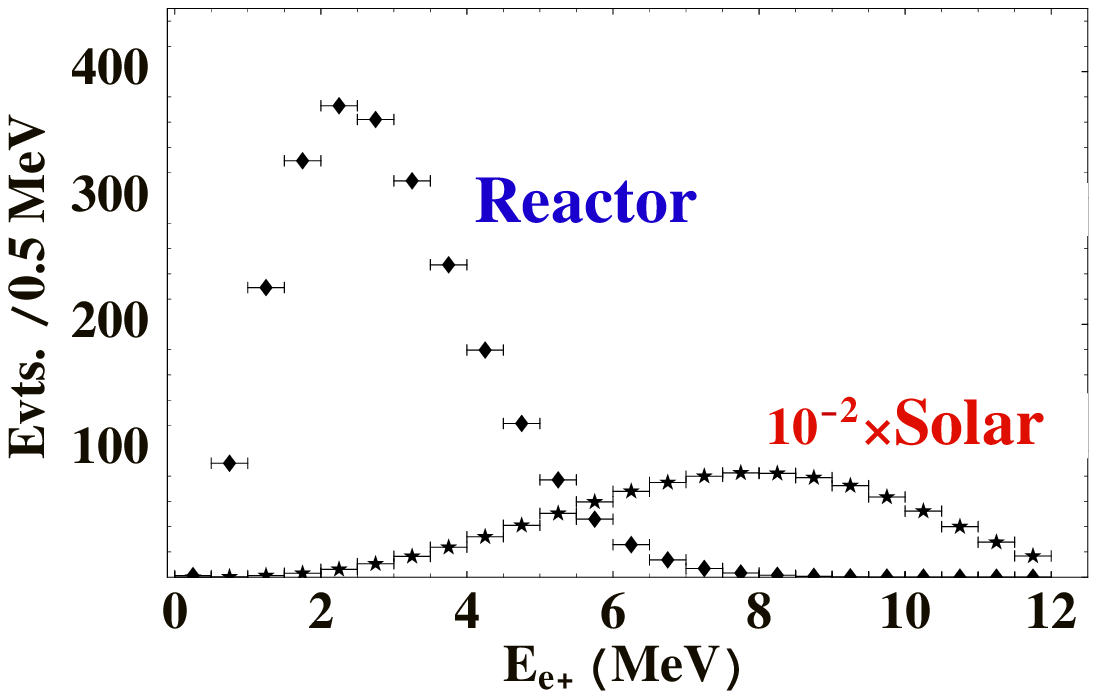,width=10cm} 
\caption{
The KamLAND expected positron spectra 
in absence of oscillations coming from reactor 
antineutrinos (normalized to three years of data taking).
The ``solar'' positron spectrum is obtained assuming the 
shape of the ${}^8 B$ 
neutrino flux and a total normalization
$10^{-2}\times \Phi({}^8 B)$ (that is, an overall 
$\nu_e-\overline{\nu}_e$ 
conversion probability $\overline{P}\sim 1\%$).}
\end{figure}


\newpage


\begin{thebibliography}{99}


\bibitem{Ahmad:2002ka}
Q.~R.~Ahmad {\it et al.}  [SNO Collaboration],
Phys.\ Rev.\ Lett.\  {\bf 89}, 011302 (2002)
[arXiv:nucl-ex/0204009].

\bibitem{Ahmad:2002jz}
Q.~R.~Ahmad {\it et al.}  [SNO Collaboration],
Phys.\ Rev.\ Lett.\  {\bf 89}, 011301 (2002)
[arXiv:nucl-ex/0204008].

\bibitem{howto} 
'How to use recent SNO results',\texttt{http://www.sno.phy.queensu.ca/} 



\bibitem{Aliani:2002ma}
P.~Aliani, V.~Antonelli, R.~Ferrari, M.~Picariello and E.~Torrente-Lujan,
arXiv:hep-ph/0205053.
P.~Aliani, V.~Antonelli, M.~Picariello and E.~Torrente-Lujan,
Nucl.\ Phys.   {\bf B634} (2002) 393-409. 
arXiv:hep-ph/0111418.
P.~Aliani, V.~Antonelli, M.~Picariello and E.~Torrente-Lujan,
Nucl.\ Phys.\ Proc.\ Suppl.\  {\bf 110}, 361 (2002)
[arXiv:hep-ph/0112101].



\bibitem{Strumia:2002rv}               
A.~Strumia, C.~Cattadori, N.~Ferrari and F.~Vissani,      
arXiv:hep-ph/0205261.  
A.~Strumia and F.~Vissani, 
JHEP {\bf 0111}, 048 (2001)[arXiv:hep-ph/0109172].



\bibitem{Bandyopadhyay:2002xj}
A.~Bandyopadhyay, S.~Choubey, S.~Goswami and D.~P.~Roy,
Phys.\ Lett.\ B {\bf 540}, 14 (2002)
[arXiv:hep-ph/0204286].



\bibitem{Barger:2002iv}
V.~Barger, D.~Marfatia, K.~Whisnant and B.~P.~Wood,
Phys.\ Lett.\ B {\bf 537}, 179 (2002)
[arXiv:hep-ph/0204253].

\bibitem{Pascoli:2002xq}
S.~Pascoli and S.~T.~Petcov,
arXiv:hep-ph/0205022.

\bibitem{deHolanda:2002pp}
P.~C.~de Holanda and A.~Y.~Smirnov,
arXiv:hep-ph/0205241.


\bibitem{Bahcall:2002hv}
J.~N.~Bahcall, M.~C.~Gonzalez-Garcia and C.~Pena-Garay,
arXiv:hep-ph/0204314.


\bibitem{Foot:2002re}
R.~Foot and R.~R.~Volkas,
arXiv:hep-ph/0204265.




\bibitem{Aliani:2002er}
P.~Aliani, V.~Antonelli, R.~Ferrari, M.~Picariello and E.~Torrente-Lujan,
arXiv:hep-ph/0206308.
E.~Torrente-Lujan,
arXiv:hep-ph/9902339.
S.~Khalil and E.~Torrente-Lujan,
J.\ Egyptian Math.\ Soc.\  {\bf 9} (2001) 91
[arXiv:hep-ph/0012203].




\bibitem{generalrandom}
E.~Torrente-Lujan,
{\it Prepared for 2nd ICRA Network Workshop: The Chaotic Universe: Theory, Observations, Computer Experiments, Rome, Italy, 1-5 Feb 1999}.
E.~Torrente-Lujan,
arXiv:hep-ph/9912225.
E.~Torrente-Lujan,
Phys.\ Rev.\ D {\bf 59}, 093006 (1999)
[arXiv:hep-ph/9807371].
E.~Torrente-Lujan,
Phys.\ Rev.\ D {\bf 59}, 073001 (1999)
[arXiv:hep-ph/9807361].
E.~Torrente-Lujan,
arXiv:hep-ph/9602398.
V.~B.~Semikoz and E.~Torrente-Lujan,
Nucl.\ Phys.\ B {\bf 556}, 353 (1999)
[arXiv:hep-ph/9809376].



\bibitem{specificrandom}
E.~Torrente-Lujan,
Phys.\ Lett.\ B {\bf 441}, 305 (1998)
[arXiv:hep-ph/9807426].





\bibitem{piepke}
A. Piepke [kamLAND collaboration], Nucl. Phys. Proc. Suppl. {\bf 91}, 99 
(2001) 
\bibitem{kamLANDmonaco}
J. Shirai, ``Start of Kamland'', talk given at {\it Neutrino 2002}, 
XXth International Conference on Neutrino Physics and Astrophysics,  
May 2002, Munich. 
Transparencies can be obtained from \texttt{http://neutrino2002.ph.tum.de}.
See also: 
P.~Alivisatos {\it et al.},
STANFORD-HEP-98-03.




\bibitem{antibounds}
E.~Torrente-Lujan,
Nucl.\ Phys.\ Proc.\ Suppl.\  {\bf 87}, 504 (2000).
E.~Torrente-Lujan,
Phys.\ Lett.\ B {\bf 494}, 255 (2000)
[arXiv:hep-ph/9911458].







\bibitem{CHOOZ}
M. Apollonio et al., CHOOZ  Coll., Phys. Lett. {\bf B 466}, 415 (1999)  
\bibitem{chooznew}
M. Apollonio et al. (CHOOZ coll.), hep-ex/9907037,\plb{466}{99}{415}.
M. Apollonio {\it et al.}, \plb{420}{98}{397}.

F.~Boehm {\it et al.},Phys.\ Rev.\  {\bf D62} (2000) 072002 [hep-ex/0003022].


\bibitem{PaloVerde}
Y.~F.~Wang  [Palo Verde Collaboration],
Int.\ J.\ Mod.\ Phys.\ A {\bf 16S1B}, 739 (2001);
F.~Boehm {\it et al.},
Phys.\ Rev.\ D {\bf 64}, 112001 (2001)
[arXiv:hep-ex/0107009].

\bibitem{brackeeler}
L.~De Braeckeleer  [KamLAND Collaboration],
Nucl.\ Phys.\ Proc.\ Suppl.\  {\bf 87} (2000) 312.
J. Shirai, ``Kamioka Liquid Scintillator Anti-Neutrino Detector'', 
Neutrino2002, May 25-30, Munich, Germany.
A.~Suzuki  [KamLAND Collaboration],
Nucl.\ Phys.\ Proc.\ Suppl.\  {\bf 77} (1999) 171.



\bibitem{usareport} J. Busenitz et al. (US KamLand Coll).
proposal for US Participation in KamLand. March 1999 (Unpublished).

\bibitem{atmospheric}
K.~S.~Hirata {\it et al.}  [Kamiokande-II Collaboration],
Phys.\ Lett.\ B {\bf 280}, 146 (1992).
T.~Toshito  [SuperKamiokande Collaboration],
arXiv:hep-ex/0105023;
Y.~Fukuda {\it et al.}  [Super-Kamiokande Collaboration],
Phys.\ Rev.\ Lett.\  {\bf 81}, 1562 (1998)
[arXiv:hep-ex/9807003];
R.~Becker-Szendy {\it et al.},
Nucl.\ Phys.\ Proc.\ Suppl.\  {\bf 38}, 331 (1995);
M.~Sanchez  [Soudan-2 Collaboration],
Int.\ J.\ Mod.\ Phys.\ A {\bf 16S1B}, 727 (2001);
M.~Ambrosio {\it et al.}  [MACRO Collaboration],
arXiv:hep-ex/0206027.






\bibitem{Aliani:2002ca}
P.~Aliani, V.~Antonelli, M.~Picariello and E.~Torrente-Lujan,
arXiv:hep-ph/0207348.
P.~Aliani, V.~Antonelli, R.~Ferrari, M.~Picariello and E.~Torrente-Lujan,
arXiv:hep-ph/0205061.


\bibitem{Barger:2001zs}
V.~Barger, D.~Marfatia and K.~Whisnant,
arXiv:hep-ph/0106207.

\bibitem{bpb2001} 
J.~N.~Bahcall, M.~H.~Pinsonneault and S.~Basu,
Astrophys.\ J.\  {\bf 555}, 990 (2001)
[arXiv:astro-ph/0010346].


\bibitem{Murayama:2000iq}
H.~Murayama and A.~Pierce,
Phys.\ Rev.\ D {\bf 65} (2002) 013012
[arXiv:hep-ph/0012075].

\bibitem{vogel}
 P.~Vogel and J.~F.~Beacom,                                      
 Phys.\ Rev.\ D {\bf 60}, 053003 (1999) [arXiv:hep-ph/9903554].  

\bibitem{zacek} G. Zacek et al. Phys. Rev. D34,9 (1986)2621.

\bibitem{reines} F. Reines, R. M. Woods, Phys. Rev. Lett. 14 (1965) 20.

\bibitem{zacek21} D. H. Wilkinson, Nucl. Phys. A377, 474 (1982).

\bibitem{PDG2002}
K. Hagiwara et al., Phys. Rev. D 66 (2002) 010001
























\end{thebibliography}
\end{document}